\documentstyle[aps,epsfig]{revtex} 
\begin{document} 
\draft
\preprint{gr-qc/0301128} 
\title{\large 
{\bf Black Hole Entropy from Spin One Punctures}}
\author{Romesh K. Kaul and S. Kalyana Rama
\footnote{email: kaul, krama@imsc.res.in}}
\address{\it The Institute of Mathematical Sciences, Chennai 600 113, India.}
\maketitle

~
%//~

\begin{abstract}
Recent suggestion, that the emission of a quantum of energy
corresponding to the asymptotic value of quasinormal modes of a
Schwarzschild black hole should be associated with the loss of
spin one punctures from the black hole horizon, fixes the
Immirzi parameter to a definite value. We show that saturating
the horizon with spin one punctures reproduces the earlier
formula for the black hole entropy, including the $ln (area)$
correction with definite coefficient $- 3/2$ for large area.
\end{abstract}
%\maketitle

One of the major achievements of the canonical quantum gravity
(or quantum geometry) framework has been confirming the
Bekenstein-Hawking area law for the entropy of large black holes
\cite{srk,aa}, originally obtained from semiclassical
considerations. It has also been argued that an area law follows
on the basis of some symmetry principles of the semiclassical
theory without any reference to detailed properties of actual
quantum states of a black hole \cite{c1}.

Another aspect of quantum black holes is the discrete spectrum
of the horizon area, first conjectured in the pioneering work of
Bekenstein \cite{b}. The canonical quantum gravity framework
does indeed predict discrete spectra for quantum area and volume
operators. The Hilbert space of canonical gravity is described
by spin networks, where SU(2) spins $(j = 0, 1/2, 1,
\cdots)$ reside on the edges of the network graph. For a surface
intersected by edges of this spin network, each puncture
carrying spin $j$ contributes an amount of area equal to $A(j) =
8 \pi l_p^2 \beta \sqrt{j (j + 1)}$ \cite{rs} where $l_p$ is the
Planck length and $\beta$ is the Immirzi parameter \cite{i}.

A detailed understanding of quantum degrees of freedom of a $(3
+ 1)-$dimensional black hole is presented in terms of an SU(2)
Chern-Simons theory on the horizon, with coupling $k$
proportional to the horizon area \cite{aa,km1,km2,dkm}.
Boundary states of a Schwarzschild black hole are charcterised
by an SU(2)$_k$ Wess-Zumino conformal field theory on this two
sphere (spatial slice of the horizon) of area $A_H$. The
dimensionality of the boundary Hilbert space then can be
calculated exactly by counting the number of conformal blocks of
this two dimensional conformal field theory with a number of
punctures produced by the edges of the spin network ending on
this two sphere. This number for a set of punctures at $p$
locations $\{1, 2, \cdots, p\}$ with spins $j_1, j_2, \cdots,
j_p$ residing on them is given by the formula \cite{km1} 
\begin{equation}\label{1}
{\cal N} = \frac{2}{k + 2} \; \sum_{r = 0}^{k/2} 
\frac{ \prod_{i = 1}^p 
sin \left( \frac{(2 j_i + 1) (2 r + 1) \pi}{k + 2} \right) }
{\left[
sin \left( \frac{(2 r + 1) \pi}{k + 2} \right) 
\right]^{p - 2}}  \; . 
\end{equation}
For a given horizon area $A_H$, the largest number of states
correspond to saturating the horizon by placing the lowest spin,
namely spin $\frac{1}{2}$, on each puncture. In reference
\cite{km2}, the formula (\ref{1}) was thus evaluated, with 
$j_i = j_{min} = \frac{1}{2}$ for all $i$, in the limit of large
$k$ and large $p$ to be
\begin{equation}\label{2} 
{\cal N} \simeq \frac{2^p}{p^{\frac{3}{2}}} \; 
\left(1 + {\cal O}\left(\frac{1}{p}\right) \right) \; .
\end{equation}
Since the associated area for each spin $\frac{1}{2}$ puncture
is $8 \pi l_p^2 \beta \sqrt{\frac{3}{4}}$, the number of
punctures for a fixed horizon area $A_H$ is given by
$p = \frac{\beta_0}{\beta} \left( \frac{A_H}{4 l_p^2} \right)$
where $\beta_0 = (\pi \sqrt{3})^{- 1}$. This when substituted in
(\ref{2}) yields for the entropy of the black hole \cite{km2}
%\begin{equation}\label{2'}
$S = ln {\cal N} = \frac{A_H}{4 l_p^2} - \frac{3}{2} \; 
ln \left( \frac{A_H}{4 l_p^2} \right) + \cdots$, ~~ 
%\end{equation}
where we have used the identification for Immirzi parameter
$\beta = \beta_0 ln 2 = \frac{ln 2}{\pi \sqrt{3}}$ to match the
coefficient of the first term to the
Bekenstein-Hawking area law. The $ln (area)$ correction term
with definite coefficient $- \frac{3}{2}$ has also been obtained
by Carlip by exploiting the nature of corrections to the Cardy
formula for the density of states in a two dimensional conformal
field theory \cite{c2}. This correction appears for a variety of
black holes independent of dimensions. In particular, the same
correction obtains for BTZ black holes in $(2 + 1)-$dimensional
theories \cite{c2,gks,bs}. As emphasised by Carlip, this
suggests a possible universal character of the $ln (area)$
correction. Same $ln (area)$ correction with coefficient 
$- 3/2$ has also been obtained by Gour \cite{g} in the 
algebraic approach of \cite{bg}. 

Recently, a case has been made by Dreyer \cite{d} that the
minimum spin value to be placed on each of the puncture should
be $j_{min} = 1$ instead of the value $\frac{1}{2}$. This
conclusion is based on a crucial observation made by Hod
\cite{h} that the real part (ringing frequencies)
$\omega^R_{QNM}$ of the (complex) quasinormal mode frequencies
$\omega_{QNM} \equiv \omega^R_{QNM} + i \omega^I_{QNM}$ for a
Schwarzschild black hole of mass $M$, earlier obtained
numerically by Nollert \cite{n} and later confirmed by Andersson
\cite{a}, has an asymptotic value 
\begin{equation}\label{3}
\omega^R_{QNM} = \frac{ln 3}{8 \pi M} 
\end{equation}
for large damping, namely for large values of the imaginary part
$\omega^I_{QNM}$ of $\omega_{QNM}$. More recently, Hod's
observation has been confirmed analytically \cite{m}. Bohr's
correspondence principle then suggests that $\omega^R_{QNM}$
should appear as a transition frequency in the quantum
theory. This in turn implies, using the area-mass relation $A_H
= 16 \pi M^2$, that the area of a black hole can change by an
amount $\Delta A_H = 32 \pi M \Delta M$ by emitting or absorbing
a quantum of energy $\Delta M = \hbar \omega^R_{QNM}$ so that
$\Delta A_H = 4 l_p^2 ln 3$ \cite{h}. On the other hand, the
area associated with a spin $j$ puncture is $8 \pi l_p^2 \beta
\sqrt{j (j + 1)}$. This, as concluded by Dreyer \cite{d},
suggests that only a spin $j = 1$ puncture is lost during the
emission of a quantum of energy $\hbar \omega^R_{QNM}$, and also
that the Immirzi parameter $\beta$ takes the value $\beta =
\frac{ln 3}{2 \pi \sqrt{2}}$ instead of the earlier preferred
value. This suggests that the underlying group for quantum
gravity is SO(3) instead of SU(2). Thus, in the calculation of
dimensionality of the boundary Hilbert space formula
(\ref{1}), which also holds for SO(3) conformal field theory
with spins restricted to integer values, we need to set $j_i =
j_{min} = 1$ on all the punctures. An immediate question to ask
is whether that changes the $ln (area)$ correction term to the
black hole entropy. We shall demonstrate here that though the
Immirzi parameter $\beta$ has a different value, yet the same
$ln (area)$ correction with definite coefficient $- {3}/{2}$
obtains even with $j_{min} = 1$ at each puncture.

To see our result, we put $j_i = j_{min} = 1$ for all $i$ 
in formula (\ref{1}) and evaluate 
\begin{equation}%\label{1}
{\cal N} = \frac{2}{k + 2} \; \sum_{r = 0}^{{k}/{2}} 
\left[ \frac{ sin \left( \frac{3 (2 r + 1) \pi}{k + 2} \right) }
{sin \left( \frac{(2 r + 1) \pi}{k + 2} \right)}  \right]^p  
sin^2 \left( \frac{(2 r + 1) \pi}{k + 2} \right) 
\end{equation}
in the limit of large $k$ and large number of punctures $p$. 
We may rewrite this expression as
\begin{equation}\label{4}
{\cal N} = 
%\frac{2}{k + 2} \; \sum_{r = 0}^{{k}/{2}} \left[1 
%+ 2 cos \left( \frac{2 (2 r + 1) \pi}{k + 2} \right) \right]^p 
%\; \frac{1}{4} \left[3 - (1 + 2 cos \left( \frac{2 (2 r + 1) 
%\pi}{k + 2} \right)) \right] 
= \frac{1}{2} \left[ 3 F(p) - F(p + 1) \right] 
\end{equation}
where we have used the formulae $sin 3 x = sin x (1 + 2 cos 2
x)$ and $4 sin^2 x = 3 - (1 + 2 cos 2 x)$, and defined 
\begin{equation}\label{5}
F(p) = \frac{1}{k + 2} \; \sum_{r = 0}^{{k}/{2}}~ \left[1 
+ 2 cos \left( \frac{2 (2 r + 1) \pi}{k + 2} \right) \right]^p ~. 
\end{equation}
This for large $k$ can be approximated by an integral
\begin{equation}\label{6}
F(p) = \frac{1}{2 \pi} \; 
\int_0^{2 \pi} d x \;  (1 + 2 cos x)^p \; . 
\end{equation}
Expanding the integral as powers of $2 cos x$ and performing 
the integration, it is straightforward to see that 
\begin{equation}\label{7}
F(p)~ =~ \sum_{l = 0}^{\left[ {p}/{2} \right] } 
~\frac{p!}{(p - 2 l)! (l!)^2} ~=~ 
\sum_{l = 0}^{\left[ {p}/{2} \right] } 
~\frac{\Gamma(p + 1)}{\Gamma(p - 2 l + 1) 
\left[ \Gamma(l + 1) \right]^2} \; . 
\end{equation}

This sum now needs to be evaluated for large $p$. This we do by
the method of steepest descent. To this effect, we write
\[
e^{f(z)}~ =~ \frac{\Gamma(p + 1)}
{\Gamma(p - 2 z + 1) ~\left[ \Gamma(z + 1) \right]^2} \; . 
\]
Thus, expanding $f(z)$ about a maximum, 
$f(z) = f(z_0) + {f''(z_0)} \; (z - z_0)^2 ~~/2 + \cdots$ 
where the maximum of $f(z)$ occurs at $z = z_0$, the method of
steepest descent yields
\[
F(p) \simeq e^{f(z_0)} \; \int dy \; 
e^{{ ~{f''(z_0)} \; y^2}~~/2} \simeq 
\sqrt{\frac{2 \pi}{- f''(z_0)}} \; ~ e^{f(z_0)} \; . 
\]
In terms of the Psi function 
$\Psi(z) = \Gamma'(z)/\Gamma(z)$, 
\begin{eqnarray*}
f'(z) & = &  ~2 \left[~ \Psi(p - 2 z + 1)~ -~ \Psi(z + 1)~ \right] \\
f''(z) & = &  ~- 2 \left[~2 \Psi'(p - 2 z + 1)~ +~ \Psi'(z + 1) 
~\right] \; . 
\end{eqnarray*}
The maximum is given by $f'(z_0) = 0$ which yields 
$z_0 = {p}/{3}$ and $f''(z_0) = - 6 \Psi'(z_0)$. Since 
$\Psi(z) \simeq ln z + \cdots$ for large $z$, one obtains  
$f''(z_0) \simeq - \frac{6}{z_0} + {\cal O}\left( \frac{1}{z_0^2}
\right) \simeq - \frac{18}{p} + {\cal O}\left( \frac{1}{p^2}
\right)$ in the limit of large $p$. $F(p)$ is then given by 
\[
F(p)~ \simeq ~\frac{\Gamma(p + 1)} {\left[ 
\Gamma(\frac{p}{3}) \right]^3} \; ~\sqrt{p}~~ \left[~ 1 
~+~ {\cal O}\left( \frac{1}{p} \right)~ \right] \; . 
\]
For large values of $z$, the Gamma function $\Gamma(z)$ 
is given by 
\[
\Gamma(z) = \sqrt{2 \pi} z^{z - \frac{1}{2}} e^{- z} 
\left[1 + \frac{1}{12 z} + \cdots \right]
\]
This yields $F(p)$ for large values of $p$:
\begin{equation}\label{8}
F(p) ~\simeq ~C' ~\frac{3^p}{\sqrt{p}}~ \left[~1~ 
-~ \frac{a}{p}~ + ~\cdots ~\right] 
\end{equation}
where $C'$ and $a$ constants independent of $p$. Thus, finally, 
the dimensionality of the boundary Hilbert space of a black hole 
for large $p$ is 
\begin{equation}\label{9}
{\cal N}~ =~\frac{1}{2}~ \left[~3 F(p)~- ~F(p + 1) ~\right] 
~= ~C ~\frac{3^p}{p^{\frac{3}{2}}}~  
\left[ ~1 ~+~ {\cal O}(\frac{1}{p}) ~\right]
\end{equation}
where $C$ is an irrelevent constant, independent of $p$. This
formula has to be contrasted with that in equation (\ref{2}) for
the case where spin $j_{min} = \frac{1}{2}$ was placed on each
puncture. The entropy of the black hole $S = ln {\cal N}$
is now 
\begin{equation}\label{10}
S~ =~ p~ ln 3~ -~ \frac{3}{2}~ ln p~ + ~{\cal O}(p^0) \; .
\end{equation}
Next, for each spin $1$ puncture, the associated area is $\Delta
A_H = 8 \pi l_p^2 \beta \sqrt{2}$, so that for black hole area
$A_H$, $p = \frac{\beta_0}{\beta} \; \left( \frac{A_H}{4 l_p^2}
\right)$, $\beta_0 = \frac{1}{2 \pi \sqrt{2}}$, and $\beta =
\beta_0 ln 3 = \frac{ln 3}{2 \pi \sqrt{2}}$.  Now, entropy
formula (\ref{10}) in terms of horison area becomes: 
\begin{equation}\label{11}
S~ = ~\frac{A_H}{4 l_p^2}~ -~ \frac{3}{2} 
~ln \left(\frac{A_H}{4 l_p^2}\right) + \cdots \; , 
\end{equation}
same as the earlier one. Clearly, though the Immirzi parameter
has changed by changing the value of  minimum spin $j_{min}$
on the punctures from ${1}/{2}$ to $1$, the entropy formula
along with its $ln (area)$ correction remains unaltered from
those found in reference \cite{km2}.

We close our discussion with a few remarks. The formula
(\ref{4}) for large $k$ only counts the number of ways SU(2)
singlets can be obtained by composing $p$ spin one
representations. To see this, we may rewrite the formula as
\[
{\cal N}~ = ~F(p)~ - ~G(p)
\]
where, using $^nC_m \equiv \frac{n!}{m! (n - m)!}$, $F(p)$ given
in equation (\ref{7}) may be rewritten as
\begin{equation}\label{13}
F(p)~ =~ \sum_{l } 
\;^pC_{2 l} ~\; \;^{2 l}C_l  
\end{equation}
and 
\begin{equation}\label{14}
G(p) ~= ~\frac{1}{2} \; \left[ ~F(p + 1)~ -~ F(p) ~\right] 
~=~ \sum_{l} ~ \frac{p!}{(p - (2 l - 1))! \; (l - 1)! \; l!} \\
~= ~\sum_{l}~ \; ^pC_{2 l - 1}~ \; ^{2 l - 1}C_l  \; .
\end{equation}
Now, $F(p)$ and $G(p)$ have a simple interpretation. $F(p)$
counts the number of ways states with net $J_3$ quantum number
$m = 0$ can be obtained by placing $m= 0$, $\pm 1$ on $p$
punctures. This can be done by picking $2 l$ preferred
punctures in $^pC_{2 l}$ ways and then place $m = + 1$ on $l$ of
them in $^{2 l}C_l$ ways and on rest $l$ punctures we place $m =
- 1$. Other $(p - 2 l)$ punctures carry $m = 0$. Then summing
over $l$ from $0$ to $\left[{p}/{2}\right]$ yields the total
number of ways ($= F(p)$) net $m = 0$ states can be obtained.
But this overcounts the number of SU(2) singlets, because net $m
= 0$ states are also contained in possible spin $j = 1, 2, 3,
\cdots, p$ representations obtained in the composition of $p$
spin $1$ representations. These extra net $m = 0$ states are
equal in number to net $m = + 1$ (or equivalently $m = - 1$)
states in spin $j = 1, 2, 3, \cdots, p$ representations of the
product. Now states with net $m = + 1$ can be obtained by
selecting $(2 l - 1)$ preferred punctures in $^pC_{2 l - 1}$
ways and placing $m = + 1$ on $l$ of them in $^{2 l - 1}C_l$
ways and $m = - 1$ on the rest $(l - 1)$ punctures. Other 
$(p - (2 l - 1))$ punctures carry $m = 0$. This is to be done
for all values of $l = 0, 1, 2, \cdots$, leading to the fact
that $G(p)$ in (\ref{14}) above counts the number of ways net 
$m = + 1$ (equivalently $m = - 1$) states can be obtained. The
difference $F(p) - G(p)$ thus does indeed count the number of
ways singlets can be obtained by composing $p$ spin $1$
representations.

In fact placing any value of spin $j_{min}$ on every puncture,
it can be shown, does not change the coefficient of the $ln
(area)$ correction for the black hole entropy. Indeed, by a
calculation similar to above, the dimensionality of the boundary 
Hilbert space of such a black hole with $j_{min}$ on every puncture
for large $p$ is
%\begin{equation}
\[
{\cal N}
~\simeq~ C ~\frac{(2 j_{min} ~+~1)^p}{p^{\frac{3}{2}}}  
\]
%\end{equation}
where $C$ is an irrelevent constant independent of $p$.

It is now clear that the resultant entropy for any value of
$j_{min}$ has the
$ln (area)$ correction term, with the definite coefficient 
$- {3}/{2}$. However, Immirzi parameter does depend on the value of
$j_{min}$. In particular, preferred value of $j_{min} = 1$
suggested by equating the mass change corresponding to the
minimum change in area $\Delta A_H = 8 \pi l_p^2 \beta
\sqrt{j_{min} (j_{min} + 1)}$ with the energy $\Delta M = \hbar
\omega^R_{QNM}$ of a quantum associated with quasinormal mode
frequency $\omega^R_{QNM}= \frac{ln 3}{8 \pi M}$, and
corresponding Immirzi parameter $\beta = \frac{ln 3}{2 \pi
\sqrt{2}}$, the entropy of a large area black hole is given by
the formula (\ref{11}), with coefficient $- {3}/{2}$ for the
$ln (area)$ correction.

~~

{\bf Acknowledgement:} Discussions with G. Date are gratefully
acknowledged.

\end{document}